\documentclass{amsart}
\usepackage{amssymb,stmaryrd}
\usepackage{amsfonts}
\usepackage{amstext}
\usepackage{algorithmic}
\usepackage{algorithm}
\usepackage{graphicx}
\usepackage{epstopdf}
\usepackage[all]{xy}

\numberwithin{equation}{section}

\theoremstyle{definition}

\theoremstyle{remark}

\newcommand{\tens}{\otimes}

\newcommand{\bDelta}{{\bar\Delta}}

\newcommand{\extd}{{\rm d}}
\newcommand{\del}{{\partial}}

\begin{document}

\title{Newtonian  gravity on quantum spacetime}
\keywords{noncommutative geometry, quantum spacetime, variable speed of light, quantum gravity}

\subjclass[2000]{Primary 81R50, 81R60, 58B32, 83C45}

\author{Shahn Majid}
\address{Queen Mary, University of London\\
School of Mathematics, Mile End Rd, London E1 4NS, UK}

\email{s.majid@qmul.ac.uk}

\date{Revised July 2012}

\begin{abstract}
The bicrossproduct model $\lambda$-Minkowski  (or `$\kappa$-Minkowski') quantum spacetime has  an anomaly for the action of the Poincar\'e quantum group which was resolved by an extra cotangent direction $\theta'$ not visible classically. We show that gauging a coefficient of $\theta'$ introduces gravity into the model. We solve and analyse the model nonrelativisticaly in a $1/r$ potential, finding an induced constant term in the effective potential energy and a weakening and separation of the effective gravitational and inertial masses as the test particle Klein-Gordon mass increases. The present work is intended as a proof of concept but the approach could be relevant to an understanding of dark energy and possibly to macroscopic quantum systems. \end{abstract}
\maketitle 

\section{Introduction} 

Quantum or noncommutative geometry\cite{Con} has been proposed for many years as a generalisation of geometry suitable to model quantum gravity corrections to classical geometry. Coming out of quantum Born reciprocity, the author proposed\cite{Ma:pla}  quantum groups as toys model with both quantum and curved phase-space. Since then many proposals have emerged for one part of that, namely flat quantum spacetimes with quantum Poincare group\cite{CW,Ma:mom,Luk,MaRue} and have led to predictions such as a  variable speed of light testable by time of flight data from gamma-ray bursts\cite{AmeMa}. There are also models \cite{DFR,Sny}  of a different character. The dual side of this is curved momentum space and was proposed by the author as a new effect called `cogravity' and related in simple cases to flat quantum spacetime by quantum Fourier transform\cite{Ma:qg}, an approach that has recently attracted some attention\cite{AFS}.  It is also now well understood in 2+1 quantum gravity how noncommutative spacetime can arise in a certain weak gravity approximation\cite{tHof,Sch,FL,FreMa} and the emergence of flat spacetimes and/or curved momentum space can be seen quite explicitly. 

In this note we propose how gravity can be included in such flat quantum spacetime models. We recall that in physics a quantum anomaly is where a classical symmetry is not preserved on quantisation. In \cite{BegMa1} we proved a no-go theorem that many classes of familiar noncommutative spaces likewise do not admit differential calculi of  classical dimensions
and which are fully covariant under expected group or quantum group symmetries. We have called this  a quantum anomaly for the differential structure and have proposed it as an algebraic origin of evolution\cite{Ma:spo}. The theorem does not specifically apply to the Poincar\'e quantum group on the Majid-Ruegg bicrossproduct model quantum spacetime
\cite{MaRue}
\begin{equation}\label{mink} [x_i,x_j]=0,\quad  [x_i,t]=\imath\lambda x_i\end{equation}
but in 2+1 this arises as a limit  of the quantum group $C_q(SU_2)$ as this is stretched flat \cite{MaSch} and there the theorem does apply. It appears that one similarly has an anomaly in all dimensions.   We will use a conventional parameter such that $\lambda\to 0$ is the classical limit rather than the original $\kappa=1/\lambda$.  

Quantum anomalies for differential structure can typically be fixed by extra cotangent directions. Thus the smallest known calculus in the 3+1 version of (\ref{mink}) is 5-dimensional and in our conventions it has the form cf\cite{Sit}
\[  [\extd x_i,x_j]=\imath\lambda\delta_{ij}\theta',\quad [\theta',x_i]=0,\quad[\theta',t]=\imath\lambda \theta'\]
\begin{equation}\label{bicrosscalc} [\extd x_i,t]=0,\quad [x_i,\extd t]=\imath\lambda \extd x_i,\quad [\extd t,t]=\beta\imath\lambda\theta'-\imath\lambda\extd t.\end{equation}
except that we have inserted a dimensionful constant $\beta$ in front of $\theta'$ for later use. The form of $\extd$ can be deduced from these relations and on normal ordered functions $\psi(x,t)=\sum_n\psi_n(x) t^n$  we have
\begin{equation}\label{extdconst} \extd \psi= {\del\over\del x_i}\psi(x, t)\extd x_i+  \del_0\psi(t)\extd t+ {\imath\lambda\over 2}\square^{\beta=const}\psi(t)\theta'\end{equation}
where 
\begin{equation}\label{waveconst}\square^{\beta=const}\psi(t)={\del^2\over\del x_i^2}\psi(t+\imath\lambda)+2\Delta^{\beta=const}_0\psi(t)\end{equation}
\[ \del_0f(t)={f(t)-f(t-\imath\lambda)\over\imath\lambda},\quad \Delta^{\beta=const}_0f(t)={\beta\over 2}\left({f(t+\imath\lambda)+f(t-\imath\lambda)-2 f(t)\over (\imath\lambda)^2}\right) .\]
Here $\square^{\beta=const}$  recovers the wave operator used on plane waves in \cite{AmeMa} to obtain the famous variable effective speed of light prediction for this model. The way that the Laplacian arises here as the `partial derivative' associated to the anomalous direction $\theta'$ is part of a `wave operator' approach to noncommutative geometry implemented in \cite{Ma:alm}. It is tied up with a deep principle of noncommutative geometry that a sufficiently noncommutative geometry is inner in the sense of a 1-form $\theta$ that generates $\extd$ by commutator and that need have no classical analogue, see \cite{Ma:spo}. In the present case $\theta=\extd t-\beta\theta'$ and in 2+1 this is a degeneration of $\theta$ for the 4D calculus\cite{Wor} on $C_q(SU_2)$.

Here $\beta=-1/c^2$ where $c$ is the classical speed of light but it turns out\cite{Ma:alm} that we still have a differential calculus for any function $\beta$. We will see that gauging this coefficient of the extra direction by allowing it to vary from point to point  introduces Newtonian gravity in the nonrelativistic limit, with $\beta$ the gravitational potential. Thus even though we work in flat spacetime its anomaly for the quantum Poincar\'e group forces an extra degree of freedom which can be viewed as the origin of gravity. We will look particularly at the $1/r$ potential for a point source at the origin. 

What we find is that the quantum mechanical limit of the Klein-Gordon equation on this noncommutative spacetime looks to first approximation as expected for a gravitational potential {\em except} that (a) there is a constant shift in the potential (b) the inertial and gravitational masses are modified, and modified differently, from the Klein-Gordon value, due to quantum-gravity corrections. Both increase as we approach and exceed the Planck scale but the gravitational one peaks and comes back down, tending towards zero. The ratio of gravitational to inertial masses peak at around 1.5 Planck masses in the model (see Figure~1). The first effect, while not yet a realistic explanation for dark energy, demonstrates a new mechanism whereby a constant energy could arise as a quantum gravity correction. 
Also, while the paper remains entirely theoretical, it is tempting to speculate that the second effect might conceivably apply to macroscopic quantum particles where if so it could be tested in an Earth based laboratory eg in \cite{RCNSC} or in the behaviour of Bose-Einstein condensates. This is speculative as it requires assuming that these objects are naturally described by a Klein-Gordon equation on the quantum spacetime with their quantum mechanical behaviour as a nonrelativistic limit, which is an unconventional point of view.

\subsection*{Acknowledgements} The present material was originally a motivational section within the preprint version of  \cite{Ma:alm} on the noncommutative black hole, but has been removed from the published version of that in favour of expansion here as a self-contained off-shoot. I also thank Nikolai Kiesel for informing me about \cite{RCNSC}.

\section{Interpretation of varying $\beta$}

When $\beta$ is not constant the  formula (\ref{extd}) continues to define the wave operator $\square$ as
\begin{equation}\label{extd} \extd \psi= {\del\over\del x_i}\psi(x, t)\extd x_i+  \del_0\psi(t)\extd t+ {\imath\lambda\over 2}\square\psi(t)\theta,'\end{equation}
 i.e. we take a point of view on the origin of the wave equation as coming out of the quantum anomaly\cite{Ma:spo,Ma:alm}. One finds that it has the form
\begin{equation}\label{wavebeta}\square\psi= \bar\Delta \psi(t+\imath\lambda)+2\Delta_0\psi,\quad \bar\Delta={\del^2\over\del x_i^2}-{1\over 2\beta}{\del\beta\over\del x_i} {\del\over\del x_i}\end{equation}
where
\begin{equation}\label{D0beta}\Delta_0 \psi(t)={\nu \psi(t+\imath\lambda)+\mu \psi(t-\imath\lambda({\beta\over\mu}-1))-(\nu+\mu)\psi(t+\imath\lambda(1-{\beta\over\nu+\mu}))\over(\imath\lambda)^2}\end{equation}
is still a `finite difference' but varying over space according to  solutions $\mu,\nu$ of the first order differential equations
\[ x_i{\del\mu\over\del x_i}+2\mu=\beta,\quad x_i{\del\nu\over\del x_i}+\nu=\mu.\]
The calculus remains locally inner with $\theta=\extd t-(\mu+\nu)\theta'$ and one still has
\[ \lim_{\imath\lambda\to 0}2\Delta_0 =\beta {\del^2\over\del t^2}\]
so that the classical limit of $\square$ is the Laplace-Beltrami operator for a metric of the static form
\begin{equation}\label{staticg}g= {1\over\beta}\extd t\tens\extd t+\extd x_i\tens\extd x_i.\end{equation}
These facts are a specialization of more general results in \cite{Ma:alm} or any Riemannian 3-manifold admitting a conformal Killing vector field, including the 3-geometry needed for the Schwarzschild black hole. 

\section{Polar coordinates in the flat spacetime bicrossproduct model}

We let $r^2=x^2$ so that $r$ is the radius from the origin. One has $r\extd r=x \extd x+\imath\lambda\theta'$
and using this there is a closed algebra of $\extd r,\theta',\extd t$ and functions of $r,t$ with \cite{Ma:alm} 
\[  [\extd r,f(r)]=\imath\lambda f'(r)\theta',\quad [\theta',f(r)]=0,\quad   [\extd r,f(t)]=0\]
\[ [f(r),t]=\imath\lambda r f'(r),\quad  [f(r),\extd t]=\imath\lambda\extd f(r),\quad r f(t)=f(t+\imath\lambda)r,\quad  \theta' f(t)=f(t+\imath\lambda)\theta'\]
and relations 
\[ [\extd t, f(t)]+\imath\lambda\extd f(t) = (\nu+\mu)\left( f(t+\imath\lambda)-f(t+\imath\lambda(1-{\beta\over\nu+\mu}))\right)\]
for any functions $f$. Here 
\[ \extd f(t)= \del_0f(t)\extd t+ {\imath\lambda}\Delta_0 f(t),\quad \extd f(r)= f'(r)\extd r+ {\imath\lambda\over 2}f''(r)\theta'\]
from the above.

The remaining commutation relations for the bicrossproduct model in polar coordinates are\cite{Ma:alm} 
\[[\extd x_i, f(r)]=\imath\lambda{x_i\over r}f'(r)\theta',\quad [\extd r,x_i]=\imath\lambda{x_i\over r}\theta',\quad [\extd x_i,{x_j\over r}]=\imath\lambda{e_{ij}\over r}\theta'\]
\[  x_i f(t)=f(t+\imath\lambda)x_i,\quad  [\extd x_i,f(t)]=0,\quad [\extd r,{x_i\over r}]=0\]
from which one can see for example that 
\[ \omega_i=\extd x_i-{x_i\over r}\extd r+\imath\lambda{x_i\over r^2}\theta',\quad [\omega_i,r]=0,\quad x_i\omega_i=0, \quad [\omega_i,x_j]=\imath\lambda e_{ij}\theta',\quad [\omega_i,t]=0. \]
Here the $\omega_i$ are the projections of the $\extd x_i$ to spheres of constant radius. Together with $\extd t,\extd r$ they cover all directions in the cotangent bundle classically and the same with $\theta'$ in the quantum case.

In the case of spherically symmetric $\beta={1\over r^n}$ one can solve the above system for $\mu,\nu$ and obtain as follows\cite{Ma:alm}:
\[ n=1:\quad\quad\quad\quad\quad\quad\quad \mu={1\over r},\quad \nu= {\ln(r)\over r},\quad  \Delta_0f(t)={1\over \imath\lambda r}({\del\over\del t}-\del_0)f(t+\imath\lambda)\quad\quad{\ }\]
\[ n=2:\quad \mu={\ln(r)\over r^2},\quad \nu={1+\ln(r)\over r^2},\quad \Delta_0 f(t)={1\over\imath\lambda  r^2}\left(\del_0 f(t+2\imath\lambda)-{\del\over\del t}f(t+\imath\lambda)\right)\]
 \[ n\ne 1,2:\quad\quad\quad\quad\quad\quad\quad\quad\quad\quad\quad  \mu={1\over (2-n)r^n},\quad \nu={1\over (2-n)(1-n) r^n}\quad\quad\quad\quad\quad\quad\quad{\ }\]
\[ \Delta_0 f(t)={1\over r^n}\left({f(t+\imath\lambda)+(1-n)f(t-\imath\lambda(1-n))-(2-n) f(t+\imath\lambda n)\over (\imath\lambda)^2(2-n)(1-n)}\right)\]
and 
\[ [\extd t, f(t)]+\imath\lambda\del_0 f(t)\extd t={1\over r^n}\left({f(t+(n-1)\imath\lambda)-f(t+\imath\lambda)\over (n-2)} \right)\theta'\]
where in the last expression the finite difference on the right is understood when $n=2$ as ${\del f(t+\imath\lambda)\over\del t}$.

\section{Reduction to Newtonian gravity}

Although Newtonian gravity does not fit exactly into general relativity, it can be modelled approximately as a metric of the form (\ref{staticg}). It is elementary to compute that for such metrics  
\[ {\rm Ricci}_{00}=\phi\bDelta^{flat} \phi,\quad \bDelta^{flat}={\del^2\over\del x_i^2},\quad \phi=\sqrt{-g_{00}}=\sqrt{-\beta^{-1}}.\]
We now suppose that 
\[ \beta=-{1\over c^2}(1-{2\Phi\over c^2})\]
where $c$ is the speed of light and for some spatially varying function $\Phi$ (the gravitational potential)  with values $<< c^2$ (a weak field approximation). So $\phi\approx c+{\Phi\over c}$ within our level of approximation and ${\rm Ricci}_{00}\approx \bDelta^{flat}\Phi$. Next, we consider an approximately static matter distribution with density $\rho$ which means stress energy tensor dominated by $T_{00}\approx \rho c^4$. Einstein's equations (in trace reversed form) read ${\rm Ricci}_{00}={8\pi G\over c^4}(T_{00}-{1\over 2}Tg_{00})$  where $T=T^\mu{}_\mu\approx-\rho c^2$ is the trace and $g_{00}=-\phi^2\approx -c^2$. Hence Einstein's equation in our approximation becomes
\[ \bDelta^{flat} \Phi=4\pi G\rho\]
as in Newtonian gravity. This is a standard derivation which we include for completeness only.

Next we consider how the associated spacetime Laplace-Beltrami wave operator changes. Classically this is 
\[  \bar\square\psi=\left(\beta{\del^2\over\del t^2}+{\del^2\over\del x_i^2}-{1\over 2\beta}{\del\beta\over\del x_i} {\del\over\del x_i} \right)\psi\approx \beta{\del^2\over\del t^2}\psi+\bDelta^{flat}\psi\]
where we can discard $-{1\over 2}\beta^{-1}\del\beta\approx \del \Phi/ c^2$ as long as the fields $\psi$ are slowly varying in space.  We {\em do not} make the same assumption about slow variation in $t$ and indeed we now consider fields of the form
\[ \psi= \Psi e^{-\imath t {m c^2\over\hbar}}\]
where $\Psi$ is slowly varying in both space and time, and where $mc^2$ is the rest mass of our test particle moving in the above geometry. In this case the spacetime wave equation $\bar\square\psi={m^2 c^2\over\hbar^2} \psi$ becomes
\[ {1\over c^2}(1-{2\Phi\over c^2})\left({m^2c^4\over\hbar^2}\Psi+2\imath {m c^2\over\hbar}\dot\Psi+\ddot\Psi\right)+ \bDelta^{flat}\Psi={m^2 c^2\over\hbar^2}\Psi\]
in which we can drop the $\ddot\Psi$ term in comparison to the others. We cancel leading terms, to obtain
\[ \imath\hbar{\del\over\del t}\Psi=-{\hbar^2\over 2m}\bDelta^{flat}\Psi+m\Phi\Psi\]
at our level of approximation, which is indeed the correct quantum mechanical description of a test particle of mass $m$ moving in a gravitational potential $\Phi$ (created by a matter density $\rho$). One can then take the classical limit of the theory to recover the classical Newtonian force of gravity. This is a different route to the one usually taken of geodesic deviation equation reducing to Newtonian motion of classical particles. 
It gives the interpretation of the parameter $\beta$ in the metric. 

\section{Effects in the quantum case}

We have looked above at the classical wave operator and its nonrelativistic limit. We now do the same for the quantum wave operator of Section~2. We are particularly interested in $\Phi=-{GM\over r}$ where $G$ is Newtons constant and $M$ is a gravitational mass concentrated at the origin and let $\gamma={2GM\over c^2}$. Then from Section~3 we have
\[ \beta=-{1\over c^2}(1+{\gamma\over r}),\quad \mu=-{1\over c^2}({1\over 2}+{\gamma\over  r}),\quad \nu=-{1\over c^2}({1\over 2}-{\gamma\over r}\ln({\gamma\over r}))\]
\[\Delta_0 f(t)=\Delta^{\beta=-1/c^2}_0 f(t)-{\gamma\over  c^2 r }\Delta_0^{hybrid}f(t+\imath\lambda),\quad  \Delta_0^{hybrid}={1\over\imath\lambda}\left({\del\over\del t}-\del_0\right)\]
We see that the effect  in $\Delta_0$ of the potential $\gamma/r$ in $\beta$ is an additional term which is a hybrid double derivative expressed as the difference of the classical and finite derivatives. 

As result, and also accounting for the term in $\bDelta$ from $\beta^{-1}\del\beta$, we have on normal ordered $\psi(x,t)=\sum \psi_n(x)t^n$ on the spacetime, 
\[ \square \psi(t)=\square^{\beta=-1/c^2}\psi(t)-{1\over 2}{\gamma\over r^3(1+{\gamma\over r})}x_i{\del\over\del x_i}\psi(t+\imath\lambda)-{2\gamma\over  c^2 r}\Delta_0^{hybrid}\psi(t+\imath\lambda)\]
as the  flat bicrossproduct spacetime wave operator (\ref{waveconst})  with correction due to the Newtonian  $\gamma/r$ potential. 
 
 In order to take a quantum mechanical limit as we did before in the classical case, we note that for any functions $f(t),g(t)$
\[ \Delta_0^{\beta=const}(fg)=(\Delta_0^{\beta=const}f)g(t+\imath\lambda)+f(t-\imath\lambda)\Delta_0^{\beta=const} g+(\del_0 f)\del_0 g(t+\imath\lambda)\]
\[
\Delta_0^{hybrid}(fg)=(\Delta_0^{hybrid}f)g+f(t-\imath\lambda)\Delta_0^{hybrid} g+(\del_0 f){\del\over\del t} g.\]
The first is a standard identity  for the finite double difference and the second proven in just the same way from the definitions. We also have to take a view on the noncommutative Klein-Gordon equation in the bicrossproduct model and we take this to be
\[ \square\psi=m^2c^2\psi.\]
In the flat space case this is justified\cite{AmeMa} by invariance under the bicrossproduct quantum Poincare group and we are making the minimum assumption that it still applies but for the wave operator quantizing the new metric (\ref{staticg}).
 
Now let normal ordered $\psi$ be of the form $\psi=\Psi(x,t)e^{-\imath{m c^2\over\hbar}t}$ with $\Psi$ slowly varying with respect to $t$ and for brevity let 
\[ \tilde m= m c^2/\hbar,\quad \zeta=e^{\tilde m\lambda}.\]
Then the noncommutative Klein-Gordon equation becomes
\begin{eqnarray*} &&\kern -20pt \zeta\bar\Delta\Psi(t+\imath\lambda)-{1\over c^2}\left( \zeta2\Delta^{\beta=1}_0\Psi+{\zeta+\zeta^{-1}-2\over(\imath\lambda)^2}\Psi(t-\imath\lambda)+2{\zeta-1\over \imath\lambda}\del_0\Psi\right)\\
&& -{\gamma\over c^2 r}\left(\zeta2\Delta_0^{hybrid}\Psi(t+\imath\lambda)+{1\over\imath\lambda}(-\imath\tilde m-{1-\zeta^{-1}\over\imath\lambda})\Psi-2\imath\tilde m\zeta\del_0\Psi(t+\imath\lambda)\right)={\tilde m^2\over c^2}\Psi.\end{eqnarray*}
We assume that $\Psi$ is slowly varying in the usual sense $|\ddot\Psi|<<\tilde m|\dot\Psi|$ of the Newtonian limit and $\lambda|\ddot\Psi|<<|\dot\Psi|$ and we assume the same for our finite difference and hybrid double time derivatives. By definition, dropping these two terms is the Newtonian limit.

We now suppose for the sake of discussion that $\lambda$ is of order the Planck time on the grounds that the noncommutativity is a quantum gravity effect. Mainly in order to simplify the equation we assume that $\Psi$ is also slowly varying compared to this time scale, so $\lambda|\ddot\Psi|<<|\dot\Psi|$ and also $\lambda|\bDelta\Psi|<<|\bDelta\Psi|$. The first means that we can approximate $\del_0\Psi\approx\dot\Psi$ while the second means that we can ignore the $t+\imath\lambda$ shift in  $\bDelta\Psi$. We also write $\Psi(t-\imath\lambda)=\Psi-\imath\lambda\del_0\Psi$. We also ignore the correction $-{1\over 2}\beta^{-1}\del\beta$ to the Laplacian as we did this in the classical analysis of the Newtonian limit. Then our equation becomes
\[ c^2\zeta\bDelta^{flat}\Psi=\left({\zeta-\zeta^{-1}\over\imath\lambda}-{\gamma\zeta\over r}2\imath\tilde m\right)\dot\Psi+\left(\tilde m^2+{\zeta+\zeta^{-1}-2\over(\imath\lambda)^2}-{\gamma\over r\imath\lambda}(\imath\tilde m+{1-\zeta^{-1}\over\imath\lambda})\right)\Psi.\]
Finally, making once again our weak field assumption that ${\gamma\over r}<<1$ we drop the ${\gamma\over r}\dot\Psi$  term 
to arrive after rearrangement at
\[ \imath\hbar{ \sinh(\tilde m\lambda)\over\tilde m\lambda}{\del\over\del t}\Psi=-{\hbar^2 e^{\tilde m\lambda}\over 2m}\bDelta^{flat} \Psi+\left(m c^2(1-{\sinh({\tilde m\lambda\over 2})\over{\tilde m\lambda\over 2}})-{G M m\over r}({\tilde m\lambda+e^{-\tilde m\lambda}-1\over{ \tilde m^2\lambda^2\over 2}})\right)\Psi\]
We have made assumptions on $\Psi$ and the field strength analogous to those that provide the Newtonian gravity limit (as explained in Section~4), hence the above should be viewed as, by definition, the exact noncommutative version of Newtonian gravity or of any other inverse square force in Newtonian mechanics (on interpreting $\gamma$ suitably). This is important because otherwise the approximations made in the derivation would typically far exceed any effects from $\lambda$. 
Working in this Newtonian gravity limit, the only assumption on $\lambda$ was with regard to $\Psi$ also slowly varying on that timescale, resulting in the finite-difference aspect of the noncommutative geometry being washed out in the approximation. This was not essential (and $\del_0$ could be used instead) but aids comparison with the usual Schroedinger picture of an inverse square force. Indeed, writing our equation in the form
\[ \imath\hbar {\del\over\del t}\Psi=-{\hbar^2\over 2 m_I}\bDelta^{flat}\Psi+ (V_0-{GMm_G\over r})\Psi\]
we see thus that the principal effects are:
\begin{enumerate} \item An effective inertial mass
\[ m_I=m { \sinh(\tilde m\lambda)\over\tilde m\lambda}e^{-\tilde m\lambda}=m(1-\tilde m\lambda+o((\tilde m\lambda)^2))\]
\item An effective passive gravitational mass
\[ m_G=m\left({\tilde m\lambda+e^{-\tilde m\lambda}-1\over{\tilde m\lambda\over 2}\sinh(\tilde m\lambda)}\right)=m(1-{\tilde m\lambda\over 3}+o((\tilde m\lambda)^2))\]
\item A constant term in the potential
\[ V_0=mc^2{\tilde m\lambda\over \sinh({\tilde m\lambda})}\left(1-{\sinh({\tilde m\lambda\over 2})\over{\tilde m\lambda\over 2}}\right)=-{mc^2\over 24}((\tilde m\lambda)^2+ o((\tilde m\lambda)^4)).\]
\end{enumerate}

These expressions are plotted in Figure 1. 
\begin{figure}
\[ \includegraphics[scale=1.2]{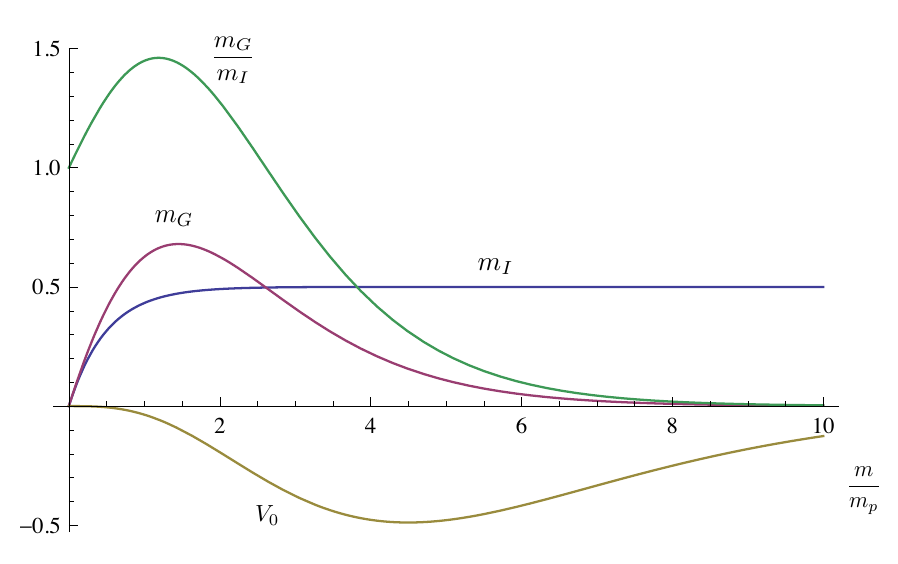}\]
\caption{Effective masses and constant energy in the model against $\tilde m\lambda=m/m_p$ where $m_p$ is the Planck mass.}
\end{figure}
The constant term $V_0$ does not arise classically but may be suggestive of some form of zero-point energy and is present even for the flat bicrossproduct model without gravity (but does not seem to have been discussed before). Its flavour is like that of dark energy in that it is present irrespective of the background matter point source, however there are also some key differences, such as the wrong sign and the fact that it is dependent on the test particle mass $m$ as something felt by the particle. On the other hand, it should be remembered that the key theoretical problem here is that conventional approaches, notably zero point energy of field oscillators in momentum space, would suggest the Planck density which is  a factor $10^{122}$  too large\cite{Dol,Wei} -- the theoretical challenge is to have a reason to make it much smaller and here we do much better due to the $\lambda^2$ in front. Thus, its value for $m=m_p$ (the Planck mass) is about -$m_pc^2/30$ and in the worst case about $-m_p c^2/2$ at about $m=4.5 m_p$. If we pretended that the universe was made up of such quantum mechanical particles spread then $V_0\sim -m_pc^2/2$  experienced by each particle in its region of space would imply a constant energy density of some
\[ - {m_p c^2\over 2}\times {m_U\over 4.5 m_p r_U^3}\]
for mass $m_U$ and radius $r_U$ of the universe (we assume some $m_U/4.5 m_p$ particles to account for the mass of the Universe). The result is something of the order of the overall energy density observed in keeping with the scale needed for dark energy (a density of about $10^{-29} g/cm^3$) in the standard cosmological model.  This does not amount to a prediction for the reasons stated, notably the negative sign. There are also conceptual issues notably should effective constant potential seen by test partlcles be seen as entering into the matter stress tensor, i.e. does it gravitate. Finally, it is not clear why we should take $m$ of order $m_P$, for normal elementary particles as test particles the effect is far smaller. Nevertheless we view the calculation   a first indication of our proposal that the cosmological constant or dark energy may have an origin as a noncommutative geometry correction, which could explain why it is so small compared to the Planck density.  

We have shown in our study of quantum black holes in \cite{Ma:alm} that due to noncommutative effects one can have standing waves inside the black hole with boundary conditions on the interior of the horizon (this is not possible classically). This supports the view that black holes do not necessarily evaporate but may form stable quantum gravity remnants where the tendency to evaporate is balanced by the need for a less massive object to have a lager Compton wavelength. Such particles would have mass of order the Planck mass as above. Their equation of state, however, would not correspond to dark energy and nor are they likely to account for dark matter  because  the rate of production of black holes in the early Universe and hence the density of possible such remnants necessarily, on energy grounds, too small. 

Turning now to $m_I$ we see that this is bounded above by $m_p/2$. This is again an effect even for the flat bicrossproduct model without gravity and is related to the well-known feature that the spatial momentum is bounded in this model. Our approach gives a new point of view on the modified dispersion relations and curved momentum space leading to the variable speed of light prediction in \cite{AmeMa}. It is striking that all the effective properties $m_I, m_G,V_0$ are bounded in the region $\pm m_p/2$ for all $m$ (the latter two go to zero as $m\to \infty$.) 

Meanwhile, we see that $m_G$ increases but more slowly than $m$ before eventually coming back down towards zero as we increase $m$. However,  $m_I$ increases even more slowly so that $m_G/m_I$ initially rises as we approach the Planck scale before peaking. Hence the effect is to make the effective gravity stronger with more acceleration of a test particle, peaking at around 1.5$m_p$ and then decaying rapidly to zero for masses $m$ much bigger than the Planck mass.

Note that usually a macroscopic object can if one wishes be treated as a limit of a quantum particle of large mass $m$, much bigger than the Planck one. It is not at all clear that this is any longer valid but if it were then we would easily be in the  paradigm covered by the above and newtonian gravity would be very far from what was observed by macroscopic objects, which is clearly not the case. However, we should remember that our analysis supposed a Klein Gordon field  with mass $m$ on the (quantum) spacetime, so we have in mind elementary particles that we might expect to be governed by such an equation, which is then seen in a quantum mechanical limit. This excludes classical macroscopic matter but possibly it does not exclude genuinely quantum macroscopic systems. It leaves open the speculative possibility that Bose-Einstein condensates and other macroscopic quantum systems might also be governed by the analysis above. This is not clear since they are effective quantum systems and as such not necessarily described by a Klein-Gordon equation, but possibly in a fully relativistic effective treatment this might be the case. 

Meanwhile, this is a speculative idea that could be put out for experimental test in the laboratory. At the moment current or proposed experiments for macroscopic quantum states are several orders of magnitude below the Planck mass but this could change. We note for example \cite{RCNSC} where it is proposed that a levitating ball be put into a quantum state and where its motion and response to external gravity can be tested. We are predicting that its response to gravity is greater than expected compared to its inertial mass as $m$ approaches the Planck scale but less and going to zero far beyond it. Note also that the effective inertial mass seems to be limited as $m$ is increased, which might translate into a different inertial mass from the mass of all the atoms in the system.

Finally it should be noted that while we have been talking about gravity, the non-constant $\beta$ could also be used to model other potentials in other contexts. Likewise the noncommutativity parameter $\lambda$ need not be the Planck time and in another context might be more easily detected.

\end{document}